\documentclass[review,12pt]{elsarticle}
\usepackage{graphicx}
\usepackage{xcolor}
\usepackage{amssymb}
\usepackage{soul}

\journal{Scripta Materialia}

\begin{document}
\begin{frontmatter}

\title{On the origin of extremely high strength of ultrafine--grained Al alloys produced by severe plastic deformation}
\author[IPAM]{R.Z. Valiev}
\author[IPAM]{N.A. Enikeev\corref{cor}}\ead{carabus@mail.rb.ru}
\author[IPAM]{M.Yu. Murashkin}
\author[IPAM]{V.U. Kazykhanov}
\author[Rouen]{X. Sauvage}
\address[IPAM]{Institute of Physics of Advanced Materials, Ufa State Aviation Technical University, K. Marx st., 12, 450000 Ufa, Russia}
\address[Rouen]{University of Rouen, Groupe de Physique des Mat\'eriaux, CNRS (UMR 6634), Avenue de l'Universit\'e --- BP 12, 76801 Saint--Etienne du Rouvray, France}
\cortext[cor]{Corresponding author}

\begin{abstract}
Ultrafine--grained Al alloys produced by high pressure torsion are found to exhibit a very high strength, considerably exceeding the Hall--Petch predictions for the ultrafine grains. The phenomena can be attributed to the unique combination of ultrafine structure and deformation--induced segregations of solute elements along grain boundaries, which may affect the emission and mobility of intragranular dislocations.
\end{abstract}


\end{frontmatter}

Grain refinement is well known to result in strength enhancement of metals and alloys, with the experimental relation between yield strength $\sigma_y$ and a mean grain size $d$ described by classic Hall--Petch relationship ~\cite{Hall, Petch}: 

$$
\label{HP}
\sigma_y = \sigma_0 + k_yd^{-1/2},
$$

where $\sigma_0, k_y > 0$ are the material's constants.  

However for nano--sized grains (20--50 nm) this relation is reported to be violated so that Hall--Petch plot deviates from linear dependence to lower stress values and its slope $k_y$ often becomes negative. In recent years this problem has been widely analyzed in both experimental and theoretical studies~\cite{PandePMS2009,LouchetPRL2006}. 

At the same time, Hall--Petch relationship breakdown is not observed in ultrafine--grained (UFG) materials with a mean grain size of 100--1000 nm usually produced by severe plastic deformation processing~\cite{ValievAlexandrovPMS2000}. Moreover, we show in this study that UFG alloys can exhibit a considerably higher strength than the Hall--Petch relationship predicts for the range of ultrafine grains. The nature of such a markedly enhanced strength is analyzed below taking into account the grain boundaries structure of UFG materials. 

The objects of this research were commercial Al alloys 1570(Al--5.7Mg--0.32Sc--0.4Mn, wt.\%) and 7475(Al--5.7Zn--2.2Mg--1.6Cu--0.25Cr, wt.\%) both having considerable content of Mg. In order to obtain UFG structure, solid--soluted alloys were subjected to high pressure torsion (HPT) at room temperature. HPT is known as one of the most effective techniques for structure refinement by severe plastic deformation (SPD)~\cite{ValievAlexandrovPMS2000}. The applied pressure of 6 GPa and number of rotations 20 were used to process the alloys. The produced samples had the form of discs with a diameter of 20 mm and 0.6 mm in thickness well suitable for mechanical tests~\cite{Murashkin1570FMM2008e}.

The structural characterization was performed by TEM, X--ray diffraction and Atom Probe Tomography (APT). A mean grain size and a grain size distribution were estimated from TEM dark field measurements in torsion plane over more than 350 grains from an area situated at the middle of an HPT disc radius. Selected area electron diffraction (SAED) patterns have been taken from an area 1.3 $\mu$m in diameter. X--ray study was performed with a Pan Analytical X'Pert diffractometer using CuK$_{\alpha}$  radiation (50 kV and 40 mA). The lattice parameter $a$ for the initial and HPT--processed alloys was calculated according to the Nelson--Riley extrapolation method \cite{KlugAlexander1974}. APT samples were prepared by standard electropolishing methods. Analyses were performed using a CAMECA Energy Compensated Atom Probe (ECOTAP) equipped with an ADLD detector~\cite{DeCostaRevSci2005}. Samples were field evaporated in UHV conditions with electric pulses (pulse fraction of 20\%, pulse repetition rate 2 kHz). The data processing was performed using the GPM 3D Data software\textcircled{R}. Tensile tests have been precisely performed using a laser extensometer at room temperature with the strain rate of $10^{-4}s^{-1}$  on computer-controlled testing machine operating with a constant displacement of the specimen grips. Strength characteristics were estimated by testing the samples with the gage of $2.0\times1.0\times0.4$ mm. 

TEM analysis proved that the HPT processing of the alloys resulted in complete refinement of the initial coarse-grained structures into an UFG ones. As an example,~Figs.~\ref{fig1tem}a,b illustrate homogeneous UFG structure formed in the HPT 1570 alloy. A grain size distribution chart, presented at~Fig.~\ref{fig1tem}c allowed to estimate a mean grain size to be equal to $\sim 97$ nm. The SAED pattern (~Fig.~\ref{fig1tem}a) exhibits typical Debye--Scherrer rings that are characteristic of ultrafine structures with mainly high angle grain boundaries. It is also important to note that a low dislocation density inside nanoscaled grains was observed in both alloys processed by HPT (Fig.~\ref{fig1tem}b), in agreement with previous studies on a Al-3\%Mg aluminium alloy processed by HPT~\cite{HoritaJMaterRes1996}.

Fig.~\ref{fig2mech} shows the results of mechanical tests of the 1570 and 7475 alloy in coarse--grained and HPT processed states. It should be noted that deformation curves for coarse--grained states are given for 1570 alloy in solid--solution state and for 7475 hardened by conventional T6 treatment. The plot demonstrates that UFG alloys manifest an outstanding strength accompanied by reduced uniform elongation. Both yield stress and ultimate tensile stress values almost three times exceed those of initial solid--soluted 1570 alloy and almost twice as higher in case of T6--treated 7475 alloy. 

Let us analyse the obtained data in terms of Hall--Petch relation to estimate to which extent the exhibited strength of UFG alloys may be determined by their grain size with special attention to 1570 alloy. There are no reference data available to construct a reliable Hall--Petch plot for the investigated alloys. In order to perform a correct comparative study we relied to the literature data for the other Al alloys.

As it is known, in case of deformed alloys a number of factors contributes to overall hardening. For the 1570 Al alloy they include hardening caused by deformation--induced structures and solid solution hardening.  That means, Hall--Petch slope for the materials processed by deformation techniques would be changed as confirmed, for example, by~\cite{TsujiNATO2006} for 1100 Al alloy. The same considerations are valid for solid solution hardening as well. 

Since the investigated UFG materials have been produced by SPD, for correct comparison we need to analyze the data obtained for Al alloys also subjected to severe straining. For that purpose two sets of data are presented (Fig.~\ref{hp}). The first one is for the 1100 Al alloy produced by accumulative roll--bonding~\cite{TsujiNATO2006} in order to outline increased $k_y$ determined by deformation--induced structures. The second one is for UFG Al--Mg alloy produced by another SPD technique --- equal--channel angular pressing~\cite{FurukawaHoritaPMA1998} and it shows simultaneous effect of both deformation and solid solution hardening. Thus, one can expect that Hall--Petch line for Al--3\%Mg alloy demonstrates slope typical for Al alloys with account to solid solution and deformation--induced hardening. 

Coarse--grained solid--solution treated 1570 Al alloy had a strength value which is somewhat higher than the strength of CG Al-3\%Mg alloy due to higher content of alloying elements, hardening contribution of which does not depend on microstructure.  TEM and XRD analysis did not reveal presence of AlMg phases in UFG 1570. One could expect that the Hall--Petch slope for 1570 would not exceed the one for Al--3\%Mg alloy. This statement can be indirectly verified by Hall--Petch data of a similar  1560 alloy (Al--6.0Mg--0.6Mn, wt.\%)~\cite{MarkushevMurashkinMSE2004}. The $k_y$ value for 1560 Hall--Petch line (Fig.~\ref{hp}, dotted line) is in a good agreement with the abovementioned suggestion. However, as one can see from Fig.~\ref{hp} the $\sigma_{02}$ value for 1570 alloy is situated significantly higher than can be predicted with respect to $\sigma_{02}$ in CG state. It means that the $\sigma_{02}$ value of 1570 alloy produced by HPT  breaks down Hall--Petch relationship approaching to higher strength. One can suppose that this phenomena could not be caused only by ultrafine grain size but also by another specific features of HPT--processed alloys microstructure. Let us analyze the nature of the observed phenomena on the example of UFG 1570 alloy via detailed examination of its microstructure features.

Thanks to XRD analysis, it was determined that HPT--processing significantly affects the crystal lattice parameter ($a$) of the Al alloys~\cite{Murashkin1570FMM2008e}. In the alloy 1570 its value is considerably reduced after processing compared to the initial state --- from $a=4.0765\pm0.0001$~\AA~to $a=4.0692\pm0.0003$~\AA.

The lattice parameter of Al--Mg alloys is directly linked to the amount of Mg in solid solution (1 at. \% Mg resulting in a change of $a$ by 0.0046~\AA~\cite{Hatch1989e}). Thus, the decrease of the lattice parameter after HPT ($\Delta a=0.0073$~\AA) can be related to loss of about 1.6 at.\% Mg by solid solution. Such a feature could be the result of deformation induced segregation or precipitation at grain boundaries. APT analyses were carried out to clarify this point. 

The amount of Mg in solid solution was measured in the annealed material (7.0$\pm$0.2 at.\%) and after HPT at room temperature (6.4$\pm$0.2 at.\%) confirming a decrease of the Mg content in solid solution. The discrepancy between the variations estimated from X-ray data ($\Delta c \approx 1.6$ at.\%) and APT analyses ($\Delta c \approx 0.6$ at.\%) might be attributed to the low statistics of APT measurements due to the small analyzed volumes. Anyway, thanks to APT the decrease of Mg content in solid solution could be attributed without any ambiguity to grain boundary segregation. As shown in the Fig.~\ref{apt}, a planar segregation of Mg was intercepted in an analyzed volume. A careful observation of the data set reveals (311) Al atomic planes on the right of the segregation, while they disappear on the left (Fig.~\ref{apt} (b)). This feature clearly indicates that there is a significant disorientation between the left and the right region separated by an excess Mg concentration along a grain boundary. Both the 2D chemical map (Fig.~\ref{apt}(c)) and the composition profile computed across the boundary (Fig.~\ref{apt}(d)) reveal that the local concentration is up to 30 at.\% Mg within a layer of about 6nm width. It is important to note that this value is much lower than the 40 at.\% expected for the Al$_3$Mg$_2$ phase, thus this Mg rich layer along the GB cannot be attributed to the intergranular precipitation of that phase. It should be noted also that Mg is not homogeneously distributed along the boundary, some significant local composition fluctuations do exist. Besides, no other elements were detected along the grain boundary.

SPD--induced grain boundary segregations in Al alloys have been already experimentally observed, for example, in UFG 6061 alloy processed by HPT~\cite{MurashkinSauvagePML2008} and in 7136 alloy processed by equal--channel angular pressing~\cite{ShaIJMR2009}. However, in these specific cases the concentration of solute elements, Mg in particular, did not exceed few atomic percents. It is natural to suppose that significantly elevated Mg content revealed in the given study could influence the mechanical behaviour of the investigated alloy. The influence should be significant, since both XRD and APT measurements testified that the Mg concentration in solid solution of UFG 1570 alloy sufficiently decreased, so one could expect certain softening of the UFG 1570 alloy instead of the observed strengthening.

It is well established that deformation of UFG materials (with grains larger than 30--50 nm) is mainly associated with intragranular movement of lattice dislocations~\cite{PandePMS2009,ValievAlexandrovPMS2000}. Besides, dislocations are generated at grain boundaries and move through a grain to be captured by an opposite grain boundary. In this case the rate--controlling mechanism is ''dislocation--grain boundary'' interaction. Elevated concentration of solutes in grain boundaries can suppress emission of dislocations from such boundaries due to solute drag. Besides, nonuniform distribution of solutes along a grain boundary would pin a dislocation  to be emitted discontinuously --- different regions of a segregation with various Mg content will drag corresponding regions of a dislocation differently, breaking it into segments at the given stress. Thus, the characteristic length, and, correspondingly, activation volume of the deformation process will be reduced.

In~\cite{ChangHongJNuclMat2008} the authors observed somewhat similar effect of sulfur segregations in cold--worked Zr--1 Nb alloy. They showed that even small amount of sulphur segregated at grain boundaries leads to noticeable decrease in $V_a$ (from 110 to 80 $b^3$ at room temperature) and additional strenthgening of the alloy. They also explained the phenomena by suggesting that sulphur atoms pin dislocation segments to be emitted from a GB. 

Let us consider the effect of reducing the activation volume in a more detailed way. The stress required for a dislocation motion can be calculated by the model of an individual dislocation emission from a grain boundary~\cite{LianAJAP2006}. According to the model, the flow stress is defined by the size of the dislocations source or the activation volume, both depending on the material structure. 

The critical stress for the emission of an individual dislocation may be expressed as~\cite{LianAJAP2006}:

\begin{equation}
\label{eqn_bow_out}
\sigma = \alpha \frac{Gb}{L}\left[ \ln \frac{L}{b} - 1.65 \right],
\end{equation}

where $\sigma$ -- yield stress, $G$ -- shear modulus, $b$ -- Burgers vector, $L$ -- the length of the dislocation or its source, $\alpha$ -- the constant. 

When deformation is realized by dislocations motion, $L$ can be expressed in terms of activation volume $v$: 

\begin{equation}
\label{eqn_ac_vol}
L = \frac{v}{b^2},
\end{equation}

Which is accordingly related to the strain rate sensitivity $m$~\cite{LianAJAP2006}: 

\begin{equation}
\label{eqn_m}
m = \frac{\sqrt{3}kT}{\sigma v}.
\end{equation} 

Based on these relations, the strain rate sensitivity can be estimated, suggesting that the increase in strength of the 1570 alloy in UFG condition is achieved due to the change in chemical composition in the grain boundary regions with corresponding change in the activation volume $v/b^3$ (or dislocation source length $L/b$). To fit the reported strength of the UFG 1570 alloy using (\ref{eqn_ac_vol}) the activation volume value should lie within the range of $v \sim 12-17 b^3$, which corresponds to $m \sim 0.02$ according to (\ref{eqn_m}). This strain rate sensitivity value is similar to the experimentally measured $m$ for the UFG Al alloys at ambient temperature~\cite{HayesActa2004} which is in agreement with the estimations of $V_a$ given above. However, to fully clarify this point, further measurements of the strain rate sensitivity of the present UFG alloy will be soon carried out.


\bibliographystyle{model1a-num-names}
\bibliography{valiev_et_al_revII}

\begin{thebibliography}{18}
\expandafter\ifx\csname natexlab\endcsname\relax\def\natexlab#1{#1}\fi
\providecommand{\bibinfo}[2]{#2}
\ifx\xfnm\relax \def\xfnm[#1]{\unskip,\space#1}\fi
\bibitem[{Hall(1951)}]{Hall}
\bibinfo{author}{E.~O. Hall}, \bibinfo{journal}{Proc. Phys. Soc. London}
  \bibinfo{volume}{64B} (\bibinfo{year}{1951}) \bibinfo{pages}{747--753}.
\bibitem[{Petch(1953)}]{Petch}
\bibinfo{author}{N.~J. Petch}, \bibinfo{journal}{J. Iron Steel Inst.}
  \bibinfo{volume}{174} (\bibinfo{year}{1953}) \bibinfo{pages}{25--28}.
\bibitem[{Pande and Cooper(2009)}]{PandePMS2009}
\bibinfo{author}{C.~Pande}, \bibinfo{author}{K.~Cooper},
  \bibinfo{journal}{Progr. Mater. Sci.} \bibinfo{volume}{54}
  (\bibinfo{year}{2009}) \bibinfo{pages}{689 -- 706}.
\bibitem[{Louchet et~al.(2006)Louchet, Weiss, and Richeton}]{LouchetPRL2006}
\bibinfo{author}{F.~Louchet}, \bibinfo{author}{J.~Weiss},
  \bibinfo{author}{T.~Richeton}, \bibinfo{journal}{Phys. Rev. Lett.}
  \bibinfo{volume}{97} (\bibinfo{year}{2006}) \bibinfo{pages}{075504(1--4)}.
\bibitem[{Valiev et~al.(2000)Valiev, Islamgaliev, and
  Alexandrov}]{ValievAlexandrovPMS2000}
\bibinfo{author}{R.~Z. Valiev}, \bibinfo{author}{R.~K. Islamgaliev},
  \bibinfo{author}{I.~V. Alexandrov}, \bibinfo{journal}{Progr. Mater. Sci}
  \bibinfo{volume}{45} (\bibinfo{year}{2000}) \bibinfo{pages}{103--189}.
\bibitem[{Murashkin et~al.(2008)Murashkin, Kilmametov, and
  Valiev}]{Murashkin1570FMM2008e}
\bibinfo{author}{M.~Murashkin}, \bibinfo{author}{A.~Kilmametov},
  \bibinfo{author}{R.~Valiev}, \bibinfo{journal}{The Phys. Met. Metallogr.}
  \bibinfo{volume}{106} (\bibinfo{year}{2008}) \bibinfo{pages}{90 -- 96}.
\bibitem[{Klug and Alexander(1974)}]{KlugAlexander1974}
\bibinfo{author}{H.~P. Klug}, \bibinfo{author}{L.~E. Alexander},
  \bibinfo{title}{X--ray Diffraction Procedures for Polycrystalline and
  Amorphous Materials}, \bibinfo{publisher}{John Wiley \& Sons},
  \bibinfo{address}{New York}, \bibinfo{year}{1974}.
\bibitem[{Costa et~al.(2005)Costa, Vurpillot, Bostel, Bouet, and
  Deconihout}]{DeCostaRevSci2005}
\bibinfo{author}{G.~D. Costa}, \bibinfo{author}{F.~Vurpillot},
  \bibinfo{author}{A.~Bostel}, \bibinfo{author}{M.~Bouet},
  \bibinfo{author}{B.~Deconihout}, \bibinfo{journal}{Rev. Sci. Instr.}
  \bibinfo{volume}{76} (\bibinfo{year}{2005})
  \bibinfo{pages}{013304--013304--8}.
\bibitem[{Horita et~al.(1996)Horita, Smith, Furukawa, Nemoto, Valiev, and
  Langdon}]{HoritaJMaterRes1996}
\bibinfo{author}{Z.~Horita}, \bibinfo{author}{D.~J. Smith},
  \bibinfo{author}{M.~Furukawa}, \bibinfo{author}{M.~Nemoto},
  \bibinfo{author}{R.~Z. Valiev}, \bibinfo{author}{T.~G. Langdon},
  \bibinfo{journal}{J. Mater. Res.} \bibinfo{volume}{11} (\bibinfo{year}{1996})
  \bibinfo{pages}{1880--1890}.
\bibitem[{Tsuji(2006)}]{TsujiNATO2006}
\bibinfo{author}{N.~Tsuji}, in: \bibinfo{editor}{Y.~T. Zhu},
  \bibinfo{editor}{V.~Varyukhin} (Eds.), \bibinfo{booktitle}{Nanostructured
  Materials by High-Pressure Severe Plastic Deformation},
  \bibinfo{publisher}{Springer Netherlands}, \bibinfo{year}{2006}, pp.
  \bibinfo{pages}{227 -- 234}.
\bibitem[{Furukawa et~al.(1998)Furukawa, Horita, Nemoto, Valiev, and
  Langdon}]{FurukawaHoritaPMA1998}
\bibinfo{author}{M.~Furukawa}, \bibinfo{author}{Z.~Horita},
  \bibinfo{author}{M.~Nemoto}, \bibinfo{author}{R.~Z. Valiev},
  \bibinfo{author}{T.~G. Langdon}, \bibinfo{journal}{Phil. Mag. A}
  \bibinfo{volume}{78} (\bibinfo{year}{1998}) \bibinfo{pages}{203--215}.
\bibitem[{Markushev and Murashkin(2004)}]{MarkushevMurashkinMSE2004}
\bibinfo{author}{M.~Markushev}, \bibinfo{author}{M.~Murashkin},
  \bibinfo{journal}{Mater. Sci. Eng. A} \bibinfo{volume}{367}
  (\bibinfo{year}{2004}) \bibinfo{pages}{234--242}.
\bibitem[{Hatch(1984)}]{Hatch1989e}
\bibinfo{editor}{J.~E. Hatch} (Ed.), \bibinfo{title}{Aluminum: Properties and
  Physical Metallurgy}, \bibinfo{publisher}{ASM Intlnl}, \bibinfo{year}{1984}.
\bibitem[{Nurislamova et~al.(2008)Nurislamova, Sauvage, Murashkin, Islamgaliev,
  and Valiev}]{MurashkinSauvagePML2008}
\bibinfo{author}{G.~Nurislamova}, \bibinfo{author}{X.~Sauvage},
  \bibinfo{author}{M.~Murashkin}, \bibinfo{author}{R.~Islamgaliev},
  \bibinfo{author}{R.~Valiev}, \bibinfo{journal}{Phil. Mag. Lett.}
  \bibinfo{volume}{88} (\bibinfo{year}{2008}) \bibinfo{pages}{459 -- 466}.
\bibitem[{Sha et~al.(2009)Sha, Ringer, Duan, and Langdon}]{ShaIJMR2009}
\bibinfo{author}{G.~Sha}, \bibinfo{author}{S.~P. Ringer},
  \bibinfo{author}{Z.~C. Duan}, \bibinfo{author}{T.~G. Langdon},
  \bibinfo{journal}{Int. J. Mater. Res.}  (\bibinfo{year}{2009})
  \bibinfo{pages}{1674--1678}.
\bibitem[{Chang and Hong(2008)}]{ChangHongJNuclMat2008}
\bibinfo{author}{K.~Chang}, \bibinfo{author}{S.~Hong}, \bibinfo{journal}{J.
  Nucl. Mater.} \bibinfo{volume}{373} (\bibinfo{year}{2008})
  \bibinfo{pages}{16--21}.
\bibitem[{Lian et~al.(2006)Lian, Gu, Jiang, and Jiang}]{LianAJAP2006}
\bibinfo{author}{J.~Lian}, \bibinfo{author}{C.~Gu}, \bibinfo{author}{Q.~Jiang},
  \bibinfo{author}{Z.~Jiang}, \bibinfo{journal}{J. Appl. Phys.}
  \bibinfo{volume}{99} (\bibinfo{year}{2006}) \bibinfo{pages}{076103(1--3)}.
\bibitem[{Hayes et~al.(2004)Hayes, Witkin, Zhou, and Lavernia}]{HayesActa2004}
\bibinfo{author}{R.~Hayes}, \bibinfo{author}{D.~Witkin},
  \bibinfo{author}{F.~Zhou}, \bibinfo{author}{E.~Lavernia},
  \bibinfo{journal}{Acta Mater.} \bibinfo{volume}{52} (\bibinfo{year}{2004})
  \bibinfo{pages}{4259--4271}.

\end{thebibliography}

\newpage
\section*{Figures captions}
\begin{figure}[t]
\begin{center}
\caption{ }
\includegraphics[angle=0, width=7.5cm]{fig1a_tem}
\includegraphics[angle=0, width=7.5cm]{fig1b_tem}
\includegraphics[angle=0, width=7.5cm]{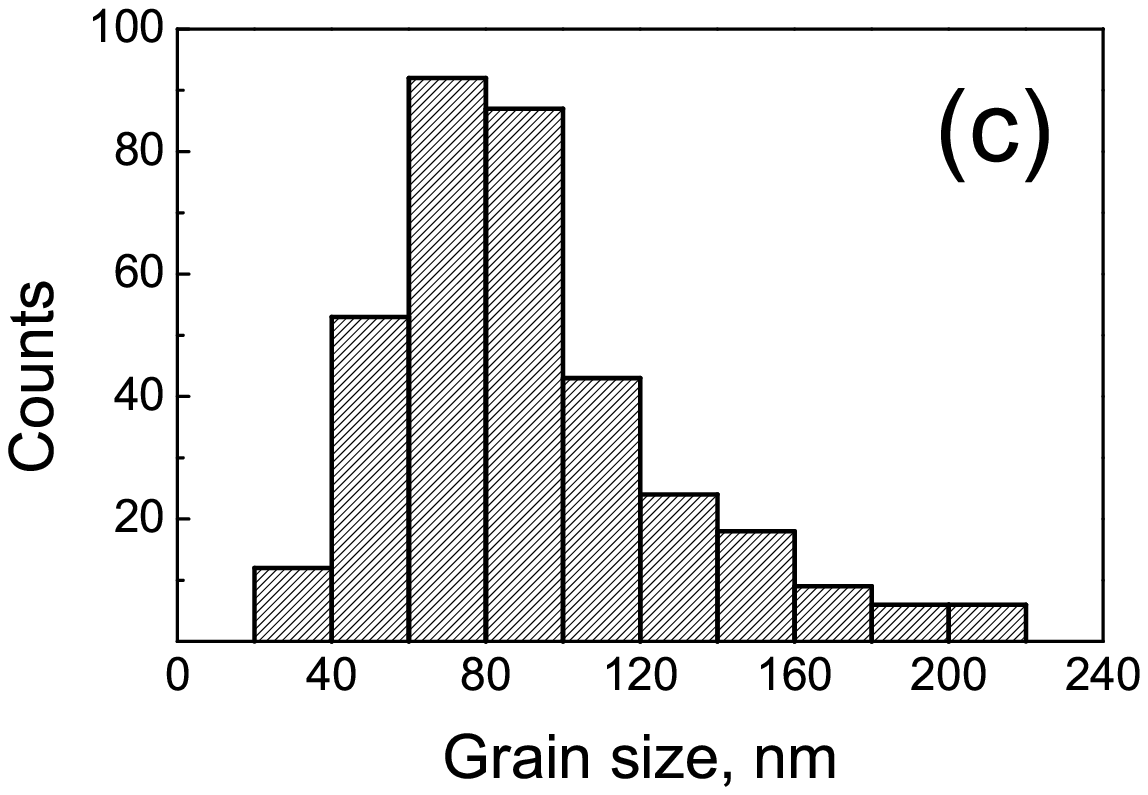}%
\label{fig1tem}
\end{center}
\end{figure}

Figure~\ref{fig1tem}. A typical TEM dark field image of the UFG 1570 alloy with a corresponding SAED (a), a bright field image (b) and a grain size distribution (c)

\begin{figure}[!ht]
\begin{center}
\caption{ }
\vspace{8pt}
\includegraphics[angle=0, width=12cm]{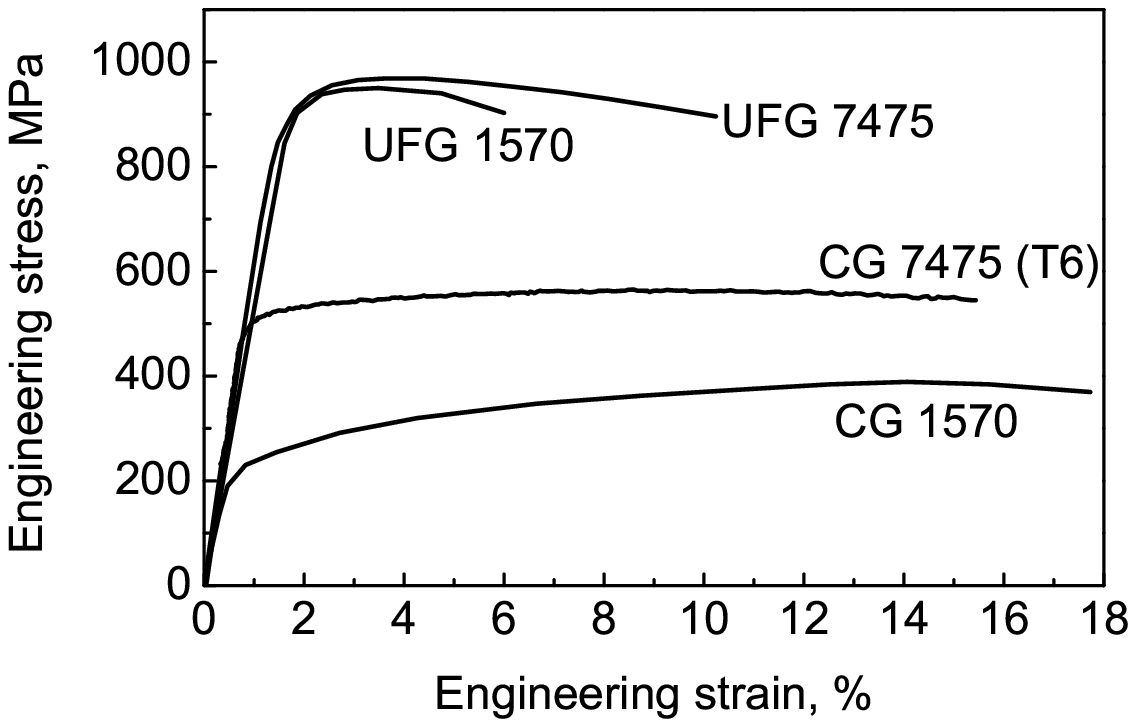}%
\label{fig2mech}
\end{center}
\end{figure}

Figure~\ref{fig2mech}. Engineering stress-strain curves for 1570 and 7475 alloys in UFG and coarse--grained states

\begin{figure}[!ht]
\begin{center}
\caption{ }
\vspace{8pt}
\includegraphics[angle=0, width=12cm]{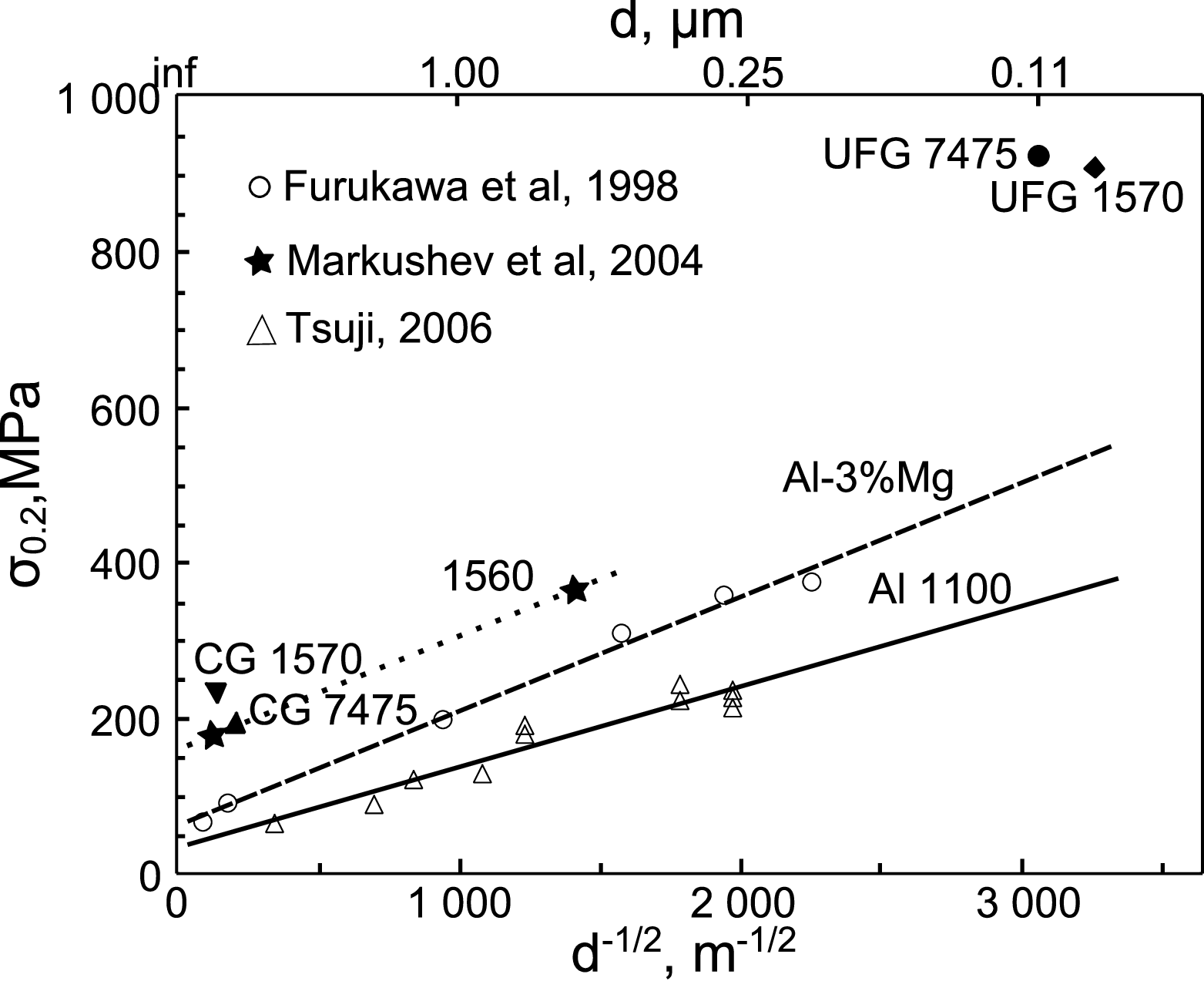}%
\label{hp}
\end{center}
\end{figure}

Figure~\ref{hp}. The Hall--Petch relation for the Al alloys: 1100~\cite{TsujiNATO2006}, Al--3\%Mg~\cite{FurukawaHoritaPMA1998} and data on the yield stresses of Al alloys: 1560~\cite{MarkushevMurashkinMSE2004}, 1570 and 7475

Figure~\ref{apt}. 3D reconstruction of an analyzed volume in the UFG 1570 alloy; (a) full data set showing a planar segregation of Mg (Al atoms are displayed in dots and Mg atoms in bubbles); (b) selected part orientated to display (311)Al atomic planes on the right of the planar segregation; (c) 2D chemical map showing the Mg concentration fluctuations within the volume; (d) concentration profile computed across the segregation (sampling volume thickness 1 nm)

\newpage

\newpage

\newpage

\newpage
\begin{figure}[!ht]
\begin{center}
\caption{ }
\vspace{8pt}
\includegraphics[angle=0, width=10cm]{fig4_v3}%
\label{apt}
\end{center}
\end{figure} 
\end{document}